\documentclass[aps,prl,twocolumn,superscriptaddress,showpacs]{revtex4-1}

\usepackage{graphicx}
\usepackage{epstopdf}
\usepackage{color}
\usepackage{epsfig}

\usepackage[english]{babel}
\usepackage{amsmath}
\usepackage{amssymb}
\usepackage{natbib}

\newcommand{\SCO}{$\rm Sr_2CuO_3$}

\begin{document}
\author{E.G.~Sergeicheva}
\affiliation{P.L. Kapitza Institute for Physical Problems, 117334 Moscow, Russian Federation}
\affiliation{Low Temperature Laboratory, Department of Applied Physics, Aalto University, P.O. Box 15100, FI-00076 AALTO, Finland}
\author{S.S.~Sosin}
\email{sosin@kapitza.ras.ru}
\affiliation{P.L. Kapitza Institute for Physical Problems, 117334 Moscow, Russian Federation}
\affiliation{National Research University Higher School of Economics, 101000 Moscow, Russia}
\author{D.I.~Gorbunov}
\affiliation{Hochfeld-Magnetlabor Dresden (HLD-EMFL) and W\"{u}rzburg-Dresden Cluster of Excellence ct.qmat,
Helmholtz-Zentrum Dresden-Rossendorf, 01328 Dresden, Germany}
\author{S.~Zherlitsyn}
\affiliation{Hochfeld-Magnetlabor Dresden (HLD-EMFL) and W\"{u}rzburg-Dresden Cluster of Excellence ct.qmat,
Helmholtz-Zentrum Dresden-Rossendorf, 01328 Dresden, Germany}
\author{G.~Gu}
\affiliation{CMPMSD, Brookhaven National Laboratory, Upton, NY 11973, USA}
\author{I.~Zaliznyak}
\email{zaliznyak@bnl.gov}
\affiliation{CMPMSD, Brookhaven National Laboratory, Upton, NY 11973, USA}

\title{Novel magnetic phase in a weakly ordered spin-1/2 chain antiferromagnet \SCO}
\date{\today}

\begin{abstract}
We present the magnetic phase diagram of a spin-1/2 chain antiferromagnet \SCO\ studied by ultrasound phase-sensitive detection technique.
We observe an enhanced effect of external magnetic field on the ordering temperature of the system, which is in the extreme proximity to the quantum critical point. Inside the N\'eel ordered phase, we detect an additional field-induced continuous phase transition, which is unexpected for a collinear Heisenberg antiferromagnet. This transition is accompanied by softening of magnetic excitation mode observed by electron-spin resonance, which can be associated with a longitudinal (amplitude) mode of the order parameter in a weakly-coupled system of spin-1/2 chains. These results suggest transition from a transverse collinear antiferromagnet to an amplitude-modulated spin density wave phase induced by magnetic field.
\end{abstract}
\maketitle

The ground state of an ideal antiferromagnetic Heisenberg spin-1/2 chain is disordered and critical, with gapless spectrum and power-law decay of spin-spin correlations \cite{Tsvelik_book}. A non-zero inter-chain interaction in material realizations of such chains leads to the formation of a long-range antiferromagnetic order at a finite temperature. When the inter-chain coupling is small, this order can be extremely weak and strongly susceptible to external perturbations, such as magnetic field. Since the properties of an ideal spin-1/2 chain are known from the exact solution \cite{Tsvelik_book}, the mean-field treatment of the exchange coupling between chains \cite{Schulz_PRL1996} allows one to obtain accurate quantitative expressions both for the ground state order parameter (the staggered magnetization, $m_0$) and the temperature of the magnetic ordering (the N\'eel temperature, $T_N$) at zero magnetic field,

\begin{equation}
m_0\approx g\mu_{\rm B} \times 1.02 \sqrt{ \frac{J_{\perp}}{J}}, ~~~~ J_{\perp} =\frac{T_N}{A\sqrt{\ln (\Lambda J/T_N)}}.
\label{1}
\end{equation}
Here, $J$ and $J_{\perp}$ are intra- and inter-chain exchange coupling constants, respectively, $g$ is a Land\'{e} spectroscopic $g$-factor, $\mu_{\rm B}$ is a Bohr magneton, and $A\approx 1.28$ and $\Lambda \approx 5.8$ are non-universal numerical coefficients.

One of the best realizations of a weakly-ordered antiferromagnetic spin-1/2 chain system known up to date is \SCO. This material has a body-centered orthorhombic crystal structure (space group $Immm$) composed of Cu-O chains running along the $b$-axis of the crystal. The strong Cu-O hybridization along this direction results in an extremely large in-chain superexchange interaction, $J\approx 2800$~K \cite{Moto_PRL1996,Suzuura_PRL1996,Walters}. The inter-chain exchange coupling is significantly reduced due to very small orbital overlaps between Cu-O ensembles in the directions perpendicular to the $b$-axis, thus resulting in an almost ideal spin-chain system. \SCO\ undergoes an ordering transition into a collinear antiferromagnetic state with ordered moments aligned with the $b$-axis at $T_N = 5.5(1)$~K ($T_N\ll J$) \cite{Keren_PRB1993,Kojima_PRL1997,SCO_PRB2017}. The exchange interaction between two nested collinear spin sublattices on a body-centered lattice is fully frustrated, so that on the mean-field level these two subsystems are exchange-decoupled. The effective inter-chain exchange coupling can be evaluated from Eq.~\eqref{1} as $J_{\perp}\approx 1.5$~K, yielding a very small ratio $J_{\perp}/J \lesssim 5\cdot 10^{-4}$. The ordered moment estimated from this ratio, $m_0\approx 0.05 \mu_{\rm B}$, is in good agreement with the experimentally observed value, $\langle\mu\rangle = 0.06\mu_{\rm B}$, so the system is in a very close proximity of the one-dimensional (1D) Luttinger-liquid quantum-critical state.

Electron-spin resonance (ESR) study \cite{SCO_PRB2017} uncovered two different types of magnetic excitations in the ordered state of \SCO. In addition to conventional pseudo-Goldstone modes with small gaps induced by weak biaxial anisotropy, a novel resonance branch was observed below $T_N$, which was related to a specific gapped longitudinal (amplitude) mode predicted by chain-mean-field (CMF) theory \cite{Schulz_PRL1996} in the proximity of a quantum-critical point. Such longitudinal mode was observed by inelastic neutron scattering in a quasi-1D spin-1/2 model system KCuF$_3$ \cite{Lake1,Lake2}. The frequency of the new ESR mode decreased with magnetic field, indicating softening at a critical field, $\mu_0 H_c\approx 9.4$~T, above which it increased again. While this observation is compatible with a second order phase transition, its presence could not be firmly established because of an incomplete gap closure.

In this work, we study the magnetic phase diagram of \SCO\ using high-sensitivity ultrasonic technique supplemented by additional ESR measurements. We observe an increase of the N\'eel temperature with magnetic field (by $\approx 20 \%$ at 15~T) consistent with the suppression of quantum fluctuations in a weakly ordered quasi-1D antiferromagnet near quantum criticality \cite{Endoh,deJonge,Birgeneau_PRB1981,Zaliznyak_SolStComm1992,ZhitomirskyZaliznyak_PRB1996}. In addition, we detect clear anomalies, both in sound velocity and in attenuation coefficient, revealing a field-induced phase transition in the field range where the additional gapped mode was found to soften. These anomalies only occur inside the ordered phase and fully disappear above $T_N$, suggesting the scenario of a continuous phase transition from the transverse N\'eel order to the longitudinal, amplitude-modulated spin density wave in magnetic field.

Our experiments were carried out on a high quality single crystal of \SCO\ ($m \simeq 0.1$~g) from the same batch as in Ref.~\onlinecite{SCO_PRB2017}. The sample was cut in a cubic shape with the edges parallel to three principal axes of the crystal lattice determined by X-ray measurements. ESR measurements on the same sample were performed to establish the equivalence of magnetic spectra to our earlier results \cite{SCO_PRB2017}. Ultrasonic measurements were carried out at Dresden High Magnetic Field Laboratory, Helmholtz-Zentrum Dresden-Rossendorf, Germany. The experimental setup was operating as a frequency variable bridge. The relative change of the sound velocity, $\Delta v/v$, was assumed to be equal to the relative change of the frequency corresponding to the bridge balance, $\Delta f/f = \Delta v/v$, neglecting the change in sample length. Ultrasound waves were generated and registered by LiNbO$_3$ single-crystal resonant piezo-electric transducers with the fundamental frequency $\simeq 27$~MHz. All the results described below were obtained for the transverse ultrasound mode propagating along $b$-axis of the crystal, $k\parallel b$, $u\parallel c$, where $k$ is the wave vector and $u$ is the polarization of the sound.

\begin{figure}[t]
\centerline{\includegraphics[width=\columnwidth]{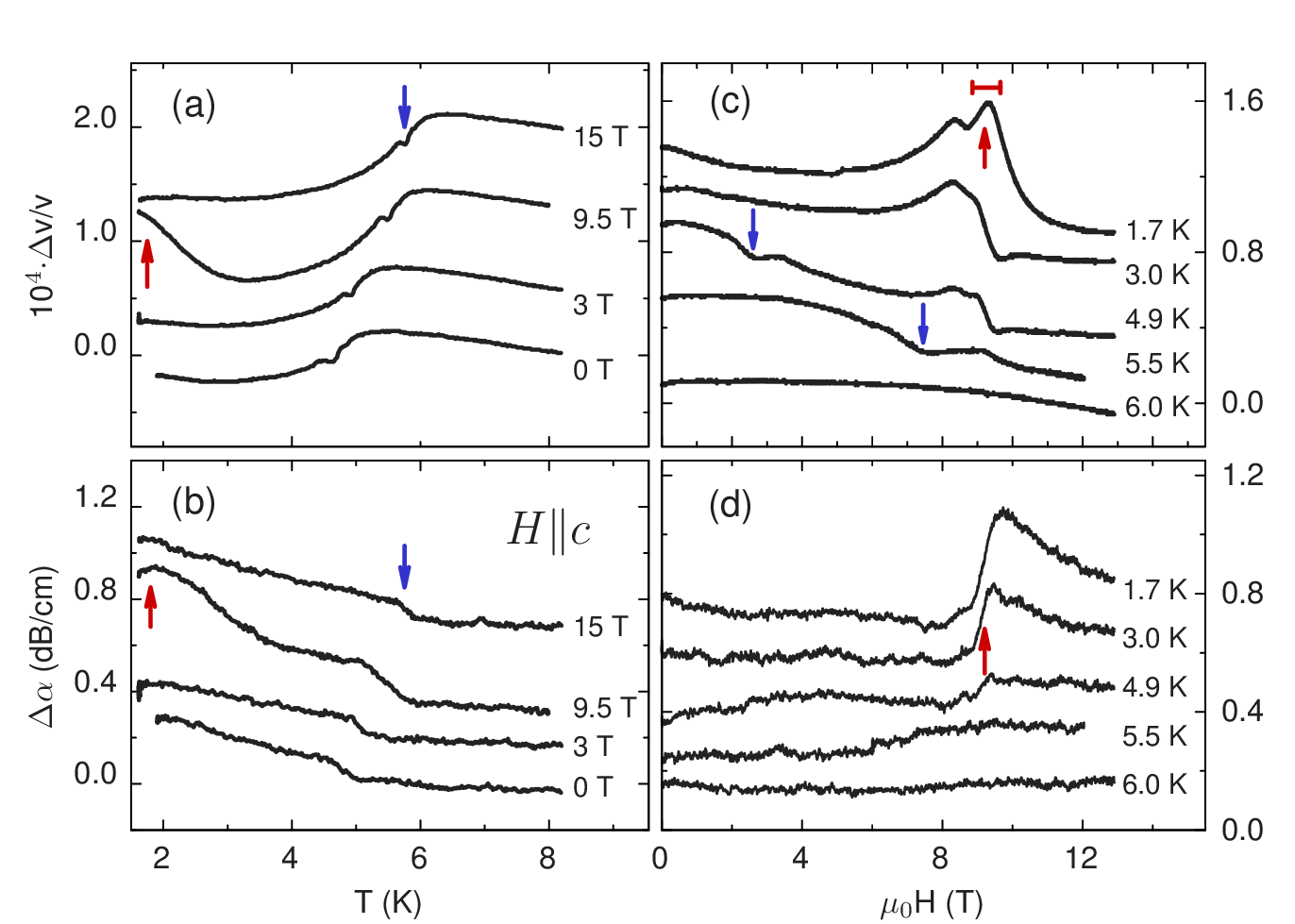}}
\caption{The temperature dependence of the relative change of sound velocity and sound attenuation of \SCO\ measured at several values of external magnetic field [panels (a,b)]; and their isothermal magnetic field dependence measured at various temperatures [panels (c,d)]. Curves, from bottom to top, are vertically shifted for clarity. The magnetic field is applied along $c$-axis, magnetic ordering and field-induced transitions are marked by down ($\downarrow $) and up ($\uparrow $) arrows, respectively.
}
\label{Hc}
\vspace{-5mm}
\end{figure}

\begin{figure}[t]
\centerline{\includegraphics[width=0.96\columnwidth]{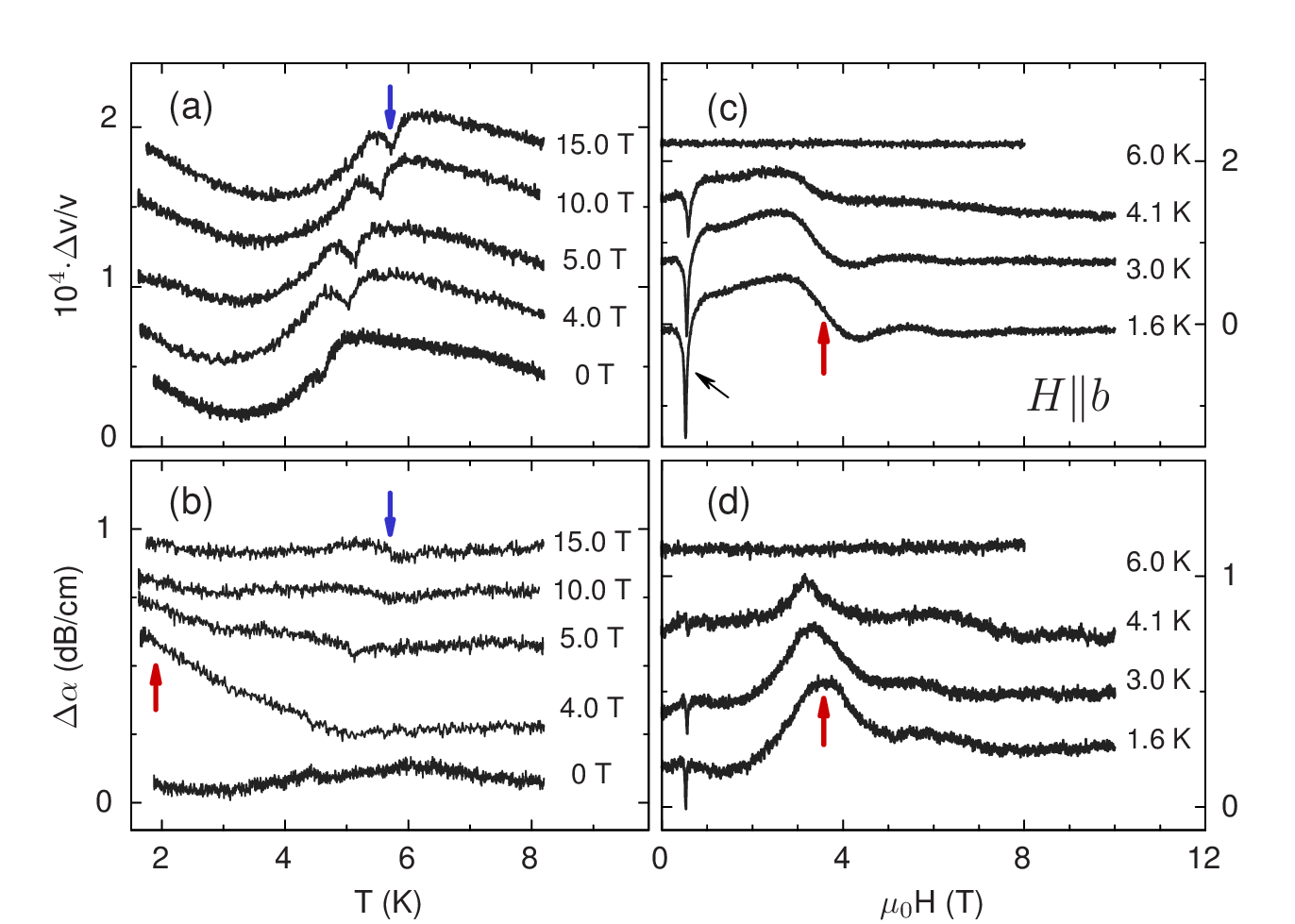}}
\caption{The relative change of sound velocity and sound attenuation measured vs temperature at several values of external magnetic field [panels (a,b)]; and vs magnetic field at various temperatures [panels (c,d)]. Curves, from bottom to top, are vertically shifted for clarity. The magnetic field is applied along $b$-axis, magnetic ordering and field-induced transitions are marked by down ($\downarrow $) and up ($\uparrow $) arrows, respectively.}
\label{Hb}
\vspace{-5mm}
\end{figure}

The temperature dependence of the relative change of the sound velocity, $\Delta v/v$, and sound attenuation, $\Delta \alpha$, in \SCO\ for different magnetic fields applied along the $c$-axis are presented in Fig.~\ref{Hc}(a),(b). The temperature dependence indicates spin-lattice coupling, because purely lattice properties are expected to be temperature independent in this range. 
The sound velocity shows slight increase with decreasing temperature, by $\sim 0.01\%$ below 15~K and down to a maximum at 5 to 6~K, depending on the field, after which it decreases. An exception is the behavior at $\mu_0 H = 9.5$~T, which is non-monotonous at low $T$, with an uprise resulting from the proximity to the $\approx 9.4$~T transition discussed below. A maximum is followed by a clear anomaly in the form of a small narrow dip in the middle of the slope, which indicates transition into a magnetically ordered state (indicated by down arrows). The transition is accompanied by a step-like increase of the sound attenuation, $\Delta \alpha$ [Fig.~\ref{Hc}(b)]. The position of the anomaly, which we associate with $T_N$, varies from about 4.7~K to 6~K, depending on the applied magnetic field.

The magnetic field dependence of $\Delta v/v$ and $\Delta \alpha$ at several constant temperatures from 1.7 to 6~K obtained by sweeping the magnetic field up to 15~T are presented in panels (c) and (d) of Fig.~\ref{Hc}. At the lowest temperature, $T=1.7$~K, the sound velocity demonstrates a relatively broad hump-like feature near $\mu_0 H \approx 9.4$~T, while the sound attenuation shows a (narrower) jump-like anomaly. The position of both features is nearly temperature independent, while they gradually decrease in amplitude with increasing temperature and completely disappear above $T_N$. The increase of the sound velocity near 9.4~T also reveals itself in the $T$-dependence of $\Delta v/v$ at 9.5~T  shown in Fig.~\ref{Hc}(a). On the other hand, the anomaly corresponding to the ordering transition at $T_N(H)$ can also be detected in the isothermal field dependences measured at 4.9~K and 5.5~K [down arrows in Fig.~\ref{Hc}(c)].

Measurements with the same sound propagation geometry were also performed for the magnetic field applied along $b$-axis. The records of the sound velocity and attenuation vs temperature appeared to be nearly identical to the previously discussed data for $H\parallel c$, with similar anomalies corresponding to antiferromagnetic transition temperature, $T_N(H)$ (panels (a) and (b) in Fig~\ref{Hb}). However, we have found a remarkable difference in the field sweeps presented in Fig.~\ref{Hb}~(c),(d). Firstly, both the sound velocity and attenuation exhibit a narrow peak at $\mu_0 H_{sf} \approx 0.5$~T, which corresponds to the spin-flop transition detected in earlier ESR-measurements \cite{SCO_PRB2017}. This peak is followed by a broader, step- or hump-like feature around 4~T, resembling the feature observed in Fig.~\ref{Hc} at $H\parallel c$ near 9.4~T. Similarly to the $H\parallel c$ case, both anomalies do not shift significantly with temperature, but decrease in amplitude and fully disappear above $T_N$.

\begin{figure}[t]
\centerline{\includegraphics[width=\columnwidth]{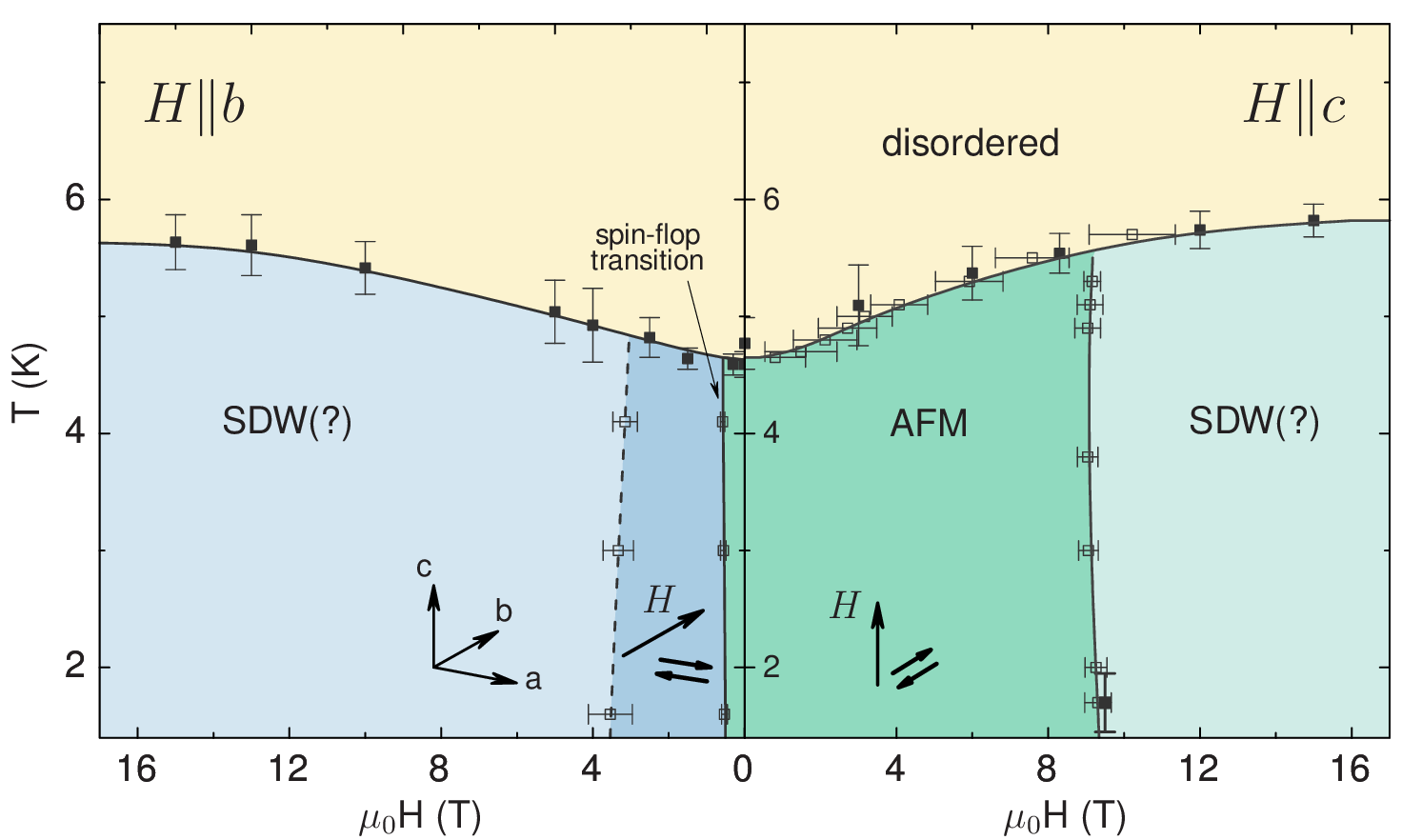}}
\caption{The magnetic phase diagram of \SCO\ obtained from ultrasound measurements at $H\parallel b$ (left panel) and $H\parallel c$ (right panel). Open and closed symbols represent features observed in field and temperature scans, respectively, and error bars represent the widths of the corresponding features. Lines are shown as guide for the eye; different magnetic phases are distinguished by colors. Arrows indicate the magnetic field and magnetic moment directions in the $a,b,c$ crystallographic coordinate system shown in the left panel.}
\label{HT}
\vspace{-5mm}
\end{figure}

Positions of the anomalies observed in all temperature and magnetic field scans are collected in Fig.~\ref{HT}, representing the magnetic phase diagram of \SCO\ for two principal directions of the applied field. The first notable peculiarity of this phase diagram is the significant increase of the temperature of magnetic ordering under applied field, which amounts to about 20\% of the zero-field $T_N$ for $\mu_0\tilde{H}\simeq 10$~T applied along $c$-axis and to about 15\% for $H\parallel b$-axis. Being in qualitative agreement with previous results for quasi-one-dimensional Heisenberg antiferromagnets (see {\it e.g.} \cite{Endoh,deJonge,Birgeneau_PRB1981,Zaliznyak_SolStComm1992,ZhitomirskyZaliznyak_PRB1996} and references therein) this finding looks exceptional from the quantitative point of view. For example, similar relative increase of $T_N$ in an organic spin-1/2 quasi-1D system CuCl$_2\cdot 2$NC$_5$H$_5$ ($J/k_B=13.4$~K, $J_{\perp}/J\sim 3\times 10^{-4}$, $T_N=1.14$~K) \cite{Endoh} is obtained under magnetic field $\tilde{H}\approx 0.35H_{sat}$, where $H_{sat}=4JS/(g\mu_B)$ is a saturation field, while the corresponding field for \SCO\ does not exceed $\tilde{H}\approx 2.5\cdot 10^{-3}H_{sat}$. Somewhat closer rate of $T_N$ variation under field ($\tilde{H}\approx 0.02H_{sat}$) was found in a nearly ideal $S=5/2$ chain system TMMC, with even smaller ratio $J_{\perp}/J=10^{-4}$. However, the effect we observe in \SCO\ is still an order of magnitude larger.

The enhanced initial uprise of $T_N$ under relatively small ($H\ll H_{sat}$) magnetic field is followed by a flattening of $T_N(H)$ curvature above 10~T, perhaps indicating the presence of an additional energy scale governing magnetic order in \SCO. Among the mechanisms contributing to the initial fast growth, in addition to magnetic field suppression of the frustration on a body centered lattice (neutron diffraction ~\cite{Kojima_PRL1997,SCO_PRB2017,Ami1995} indicates degenerate collinear structures in \SCO\ below $T_N$), could be a presence of a tiny amount of impurities \cite{SCO_PRB2017}. According to NMR experiments \cite{takigawa1997}, magnetic field can induce local areas of staggered magnetization around impurities, which extend with decreasing temperature thus possibly stimulating magnetic ordering in a system of weakly coupled spin chains. The observed slowing down of $T_N(H)$ dependence, however, is most likely related to the transformation of the spin ground state revealed by the field-induced transition reported here.

For the magnetic field applied along $c$-axis of the crystal, the field range of rapid variation of $T_N$ vs $H$ ends with the field-induced phase transition at $\mu_0H = 9.3\pm 0.2$~T. The value of the critical field is nearly temperature independent up to the transition into the paramagnetic phase at $T_N(H)$ (Fig.~\ref{HT}, right panel). For $H\parallel b$, the similar features on the field dependencies of sound velocity and attenuation indicating the transition are observed at lower fields, between 3 and 4~T (Fig.~\ref{HT}, left panel).

\begin{figure}
\centerline{\includegraphics[width=0.8\columnwidth]{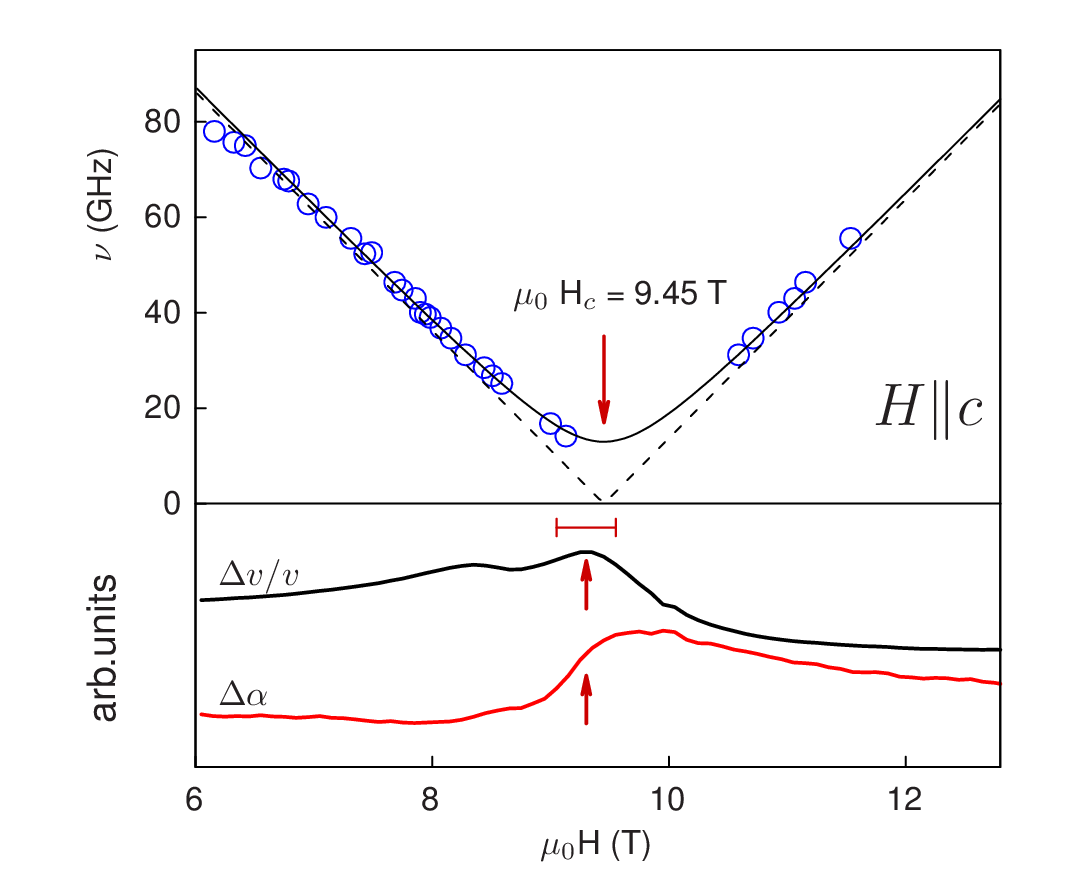}}
\caption{High-field part of the frequency-field diagram of the magnetic resonance spectrum in the ordered phase of \SCO\ measured at $T=1.5$~K for \mbox{$H\parallel c$}; symbols are experimental points including data from~\protect\cite{SCO_PRB2017}, solid line is a fit to equation $h\nu=\sqrt{(h\delta )^2 +(g_{eff}\mu_B)^2(H-H_c)^2}$ with the critical field $\mu_0H_c=9.45$~T (marked by arrow) and the residual gap $\delta =13$~GHz. Dashed line corresponds to a critical-type linear dependence with $\delta =0$. Lower panel shows the lowest temperature curves for $\Delta v/v$ and $\Delta\alpha$ from Fig.~\ref{Hc}; the anomalies at $H_c$ are pointed out by arrows and their width is shown by a horizontal bar.}
\label{Resonance}
\vspace{-5mm}
\end{figure}

The ESR measurements presented in Fig.~\ref{Resonance} (upper panel) show that the transition at $H\parallel c$ coincides with the softening of the unusual spin excitation mode, which was discovered in our recent ESR experiments and interpreted as the coupled mode of Goldstone spin wave and the longitudinal (amplitude) mode of the order parameter~\cite{SCO_PRB2017} (no similarly softening resonance mode is seen at $H\parallel b$ in that frequency range). The observed non-monotonic field dependence can be described by the critical behavior of the resonance frequency, $\nu \propto |H-H_c|$, expected for a continuous second order phase transition (Fig.~\ref{Resonance}). A small residual gap introduced to best fit the data can be explained by the mode-repulsive coupling of this excitation with the field-independent pseudo-Goldstone magnon at $\nu \approx 13$~GHz \cite{SCO_PRB2017}.
Alternatively, this gap might reflect a slight (by $1 - 2^{\circ}$) deviation of the applied magnetic field from $c$-axis of the sample.

Extrapolating the field dependence of the softening mode to $H = 0$, one can estimate its zero-field energy, $h\nu (H=0)\simeq 0.7\pm 0.1$~meV. This value compares very favorably with the energy of a longitudinal mode predicted by chain mean field theory for weakly coupled spin-1/2 chains in the proximity of the critical point, $\varepsilon_L = \sqrt{2/3}\Delta \approx 0.65$~meV, where $\Delta\approx 6.2J_{\perp}\simeq 0.8$~meV is a ``mass gap'' \cite{Schulz_PRL1996}. Inelastic neutron scattering experiments in another $S=1/2$ chain material, KCuF$_3$, provided evidence for the existence of such mode, albeit heavily damped \cite{Lake1}, which hinders its experimental observation by means of magnetic resonance. In \SCO\ the softening mode is reasonably sharp affording its detection by ESR, which is probably due to the remarkable one-dimensionality of this material. Attributing this mode to the longitudinal (amplitude) mode of the order parameter, one can identify the field-induced phase transition accompanied by its softening as a symmetry breaking transition to the amplitude-modulated, longitudinal spin density wave (LSDW) state. An LSDW phase induced by magnetic field was previously observed in a system of weakly coupled Ising-like, spin-1/2 XXZ chains \cite{Grenier_PRB2015}, but not for the Heisenberg case studied here.

In conclusion, ultrasound measurements in \SCO\ single crystal were used to obtain the $(H,T)$ magnetic phase diagram in a wide range of temperatures and magnetic fields applied along two principal axes. The results reveal clear anomalies in the ultrasound velocity and attenuation, which identify the temperature of magnetic ordering transition, $T_N$, the spin-flop, and the novel field-induced magnetic transition inside the ordered phase. The value of a zero-field N\'eel temperature $T_N(H=0)\simeq 5$~K has been found to increase markedly with magnetic field, by about 20\% at 10~T. The enhanced suppression of quantum fluctuations by a relatively weak magnetic field, $g\mu_{\rm B}H\ll 4JS$, should be related to the proximity of the system to the quantum critical point. Our central result is the novel field-induced second order magnetic transition at $g\mu_{\rm B}H_c\ll J$. The critical field of this transition depends on magnetic field orientation but was found to be nearly temperature-independent, with the magnitude of the corresponding anomalies vanishing above $T_N$. The ESR measurements show that this transition is accompanied by a linear softening and reopening of an unusual gapped mode of magnetic excitation, which can be associated with the longitudinal (amplitude) mode of the order parameter. This suggests field-induced transformation of the initial collinear antiferromagnet into an amplitude-modulated LSDW state at $g\mu_{\rm B}H_c\ll J$, which is surprising \cite{OkunishiSuzuki_PRB2007} in the case of weakly coupled nearly-isotropic Heisenberg chains in Sr$_2$CuO$_3$ studied here.

The authors are grateful to A.~I.~Smirnov, L.~E.~Svistov, and M.~Takigawa for valuable discussions. This research has been supported by the Russian Science Foundation, grant No.~17-12-01505 and by the Basic research Program of HSE. We acknowledge support of the HLD at HZDR, member of the European Magnetic Field Laboratory (EMFL), the DFG through SFB 1143 and the excellence cluster ct.qmat (EXC 2147, Project ID 39085490). The work at Brookhaven was supported by the Office of Basic Energy Sciences, U.S. Department of Energy (DOE) under Contract No. DE-SC0012704.

\end{document}